\documentclass[aps,pre,showpacs,floatfix,preprint]{revtex4}
\usepackage{amsmath,amssymb,color,graphicx,array,bm,url}

\begin{document}

\title{Spatial dimensionality affects the convective thawing of permafrost}

\author{M. Magnani$^{1,2}$}
\author{S. Musacchio$^{2,3}$}
\author{G. Boffetta$^{2,3}$}
\thanks{Corresponding author}
\email{marta.magnani@unito.it}
\affiliation{$^1$Institute of Atmospheric Sciences and Climate, National Research Council (CNR-ISAC), Corso Fiume 4, 10133 Torino, Italy}
\affiliation{$^2$INFN, Sezione di Torino, via P. Giuria 1, 10125 Torino, Italy.}
\affiliation{$^3$Dipartimento di Fisica, Universit\`a degli Studi di Torino, via P. Giuria 1, 10125 Torino, Italy.}

\begin{abstract}
Permafrost thaw is one of the major consequences of ongoing climate change, 
with potentially far-reaching implications for high-latitude and high-altitude environments. 
Because permafrost degradation is a subsurface phenomenon, its dynamics 
and future evolution are primarily investigated through numerical models. 
Yet, the role of groundwater convection in 
permafrost thawing is an underrepresented process. 
Here, we explore how the spatial dimensionality affects the simulation of permafrost 
thaw and heat transport within the thawed layer, known as the active layer. 
Using two- and three-dimensional (2D and 3D) simulations, 
with realistic parameter sets that enable convection, 
we show that permafrost thawing rate and heat transport are systematically 
slower in 2D than in 3D simulations. The ice-water interface develops spatial
patterns that coincides with the underlying convective structures, 
with rising cold plumes and descending warm plumes shaping its morphology. 
These findings reveal a coupling between convective flow structures and 
the evolving geometry of the permafrost boundary, and highlight the importance 
of three-dimensional representations of subsurface heat transport for accurately 
capturing permafrost thaw dynamics in land-surface models.
\end{abstract}

\date{\today}

\maketitle

\section{Introduction} 

Permafrost evolution is governed by thermal and hydrological processes 
occurring at the ground surface and within the active layer, i.e. 
the near-surface layer that undergoes seasonal thawing and
freezing, typically overlying permafrost.
However, the harsh environmental conditions of permafrost regions make 
field observations challenging, resulting in datasets that remain sparse in 
both spatial and temporal coverage \cite{karjalainen2019,obu2019}.
Thus, numerical models are essential tools for investigating 
permafrost behaviour and understanding active-layer dynamics 
\cite{chadburn2017,obu2019}.

Land-surface models generally project permafrost dynamics by considering vertical
conductive heat transfer through the active layer
\cite{dankers2011,wania2009,guimberteau2018,lawrence2008,andresen2020}.
Yet, fluid convection within the active-layer soil matrix may also
substantially influence the rate of permafrost thaw, particularly in highly
permeable soils \cite{magnani2024convection}. 
This process is driven by the anomalous behavior of
water density. During thawing, temperature within the active layer decreases with
depth, from the warm ground surface to the freezing temperature at the
permafrost surface \cite{biskaborn2019permafrost}. 
Along this thermal gradient, water density varies
nonlinearly: below approximately $4~^\circ$C, water exhibits a negative thermal
expansion coefficient, meaning its density increases with temperature, whereas
above this threshold the relationship reverses. Consequently, in the lower
portion of the active layer, where $0~^\circ$C $< T < 4~^\circ$C, pore water
becomes gravitationally unstable, creating conditions favorable for convective
flow.

Explicitly resolving convective processes in land-surface models can be
computationally expensive, making parameterisations a practical alternative for
representing their effects without simulating convection directly. 
Developing accurate parameterisations, however, requires systematic investigation
of different combinations of soil and thermal properties through dedicated studies
(e.g. \cite{magnani2024convection}).
Running two-dimensional (2D) simulations instead of three-dimensional (3D)
ones could substantially reduce computational costs, thereby facilitating this
task.  

In this work we study numerically the dynamics of the active layer and the thawing
of permafrost, comparing 2D and 3D simulations with analogue 
parameter sets. 
We consider an homogeneous porous medium saturated with water
in the presence of gravity. 
The initial configuration for both 2D and 3D simulations consists of a thawed
upper layer, at initial temperature $T_1>T_M=0~^\circ$C,
above a lower frozen layer at temperature $T_0<T_M$. 
We choose realistic parameter values and investigate three 
soil permeability values that allow convective motion to develop.

We find that the process of permafrost thaw is faster in three than in two dimensions. 
In the 3D simulations, the heat transport is stronger and the modulation of the 
water-ice interface caused by convective motions is less self-correlated in time 
compared to 2D simulations. Such evidences highlight the critical role of dimensionality 
in the simulation of convection within the active layer. 
 


\section{Physical model and numerical method}

The physics of thawing in a porous medium is described by a phase field 
method, a convenient analytical and numerical tool for 
simulating the dynamics of multiphase fluids in the presence of 
phase transitions 
\cite{beckermann1999modeling,favier2019rayleigh,yang2022abrupt}.
The physical state of water is represented by a continuous phase field
$\phi({\bm x},t)$ which takes the value $\phi=0$ in the liquid phase 
and $\phi=1$ in the solid phase. The equation for the phase
field is a Allen-Cahn type equation \cite{hester2020improved} coupled with 
an advection-diffusion equation for the temperature
field $T({\bf x},t)$ and the Darcy's equation, which describes the 
velocity field ${\bf u}({\bf x},t)$ in the porous medium 

\begin{equation}
{\bm u} = {k \over \mu \varphi} (-{\bm \nabla}p+(1-\phi)^2 {\bm g} \rho)
\label{eq1}
\end{equation}
\begin{equation}
{\partial T \over \partial t} + {\bm u} \cdot {\bm \nabla} T =
\kappa \nabla^2 T + \Delta T St
{\partial \phi \over \partial t}
\label{eq2}
\end{equation}
\begin{equation}
{\partial \phi \over \partial t} =
{6 \kappa' \over 5 St} \left[ \nabla^2 \phi -
{1 \over \delta^2} \phi (1-\phi)\left(1 - 2 \phi + 
{T - T_M \over \Delta T}\right) \right]
\label{eq3}
\end{equation}
where ${\bf g}=(0,0,-g)$ represents gravity, $k$ is the permeability,
$\mu$ is the fluid viscosity, $\varphi$ the porosity, $p$ is the 
total pressure and $\kappa$ represents an effective thermal diffusivity 
(linear combination 
of the solid and the liquid contributions \cite{rees2000vertical}).
$St=L/(c_p \Delta T)$ is the dimensionless Stefan number 
defined in terms of the latent heat for unit mass $L$, the specific 
heat capacity of water $c_p$ and the temperature jump $\Delta T=T_1-T_0$,
while $\delta$ and $\kappa'$ 
represent the interface thickness and mobility.
$T_M=0~^{o}C$ is the melting temperature.
Finally, density and temperature of water are related by the
empirical non-monotonic model, valid around the temperature 
$T_{max}=3.98 \, ^oC$ of maximal density, 
$\rho=\rho_0 (1 - \alpha^* \lvert T - T_{max} \rvert ^q )$ 
where $\alpha^*=9.3 \times 10^{-6} \, K^{-q}$ 
is a generalized thermal expansion coefficient,
$q =1.895$ and $\rho_0 = 999.97  \, kg \, m^{-3}$ \cite{wang2021growth}.
We remark that the model (\ref{eq1}-\ref{eq3}) assumes 
local thermal equilibrium by which a single 
temperature field describes the solid and the fluid phases
in the porous medium \cite{lapwood1948convection}. 
More complex, non-equilibrium, models 
which take into account heat exchanges within the 
solid matrix have been proposed
\cite{rees2000vertical,banu2002onset} and could be used for future studies in 
the present setup.

For simulating the dynamics of the active layer, we consider an initial 
configuration of an upper thawed layer of depth $H(t=0)=H_0$  
above a frozen region representing the permafrost.
The ground surface has a constant temperature $T_1 > T_M$ 
and the bulk permafrost has a uniform temperature $T_0 < T_M$. 
The upper soil layer is initially set to a linear temperature profile
$T(z) = T_1+(T_1-T_M)z/H_0$ corresponding to a diffusive solution,
which connects the ground-surface temperature $T_1$ at $z=0$
to the temperature $T_0$ of the permafrost bulk at $z_b = H_0(T_0-T_1)/(T_1-T_M)$.    
Because of the density anomaly of water, the thawed layer where 
$T_M < T < T_{max}$ has an unstable stratification and can develop
convective motions.
These are controlled by the the Rayleigh-Darcy number
defined as 

\begin{equation}
Ra = {g \Delta \rho k H \over \varphi \mu \kappa} 
\label{eq4}
\end{equation}
where $H$ is the thickness of the thawed layer with unstable stratification
and $\Delta \rho = \rho(T_{max})-\rho(T_M)$ is the density difference.
We remark that (\ref{eq4}) can be rewritten as $Ra=U_0 H/\kappa$ 
where $U_0=g \Delta \rho k/(\varphi \mu)$ is the characteristic
velocity defined by (\ref{eq1}). 
For values of $Ra$ below the critical value $Ra_c=4 \pi^2$, 
the diffusive solution with ${\bf u}=0$ is stable.
In contrast, when $Ra>Ra_c$ the unstable density stratification
is expected to develop convective motion \cite{nield2006convection}.
When $U_0=0$ (and thus $Ra=0$) the model (\ref{eq1}-\ref{eq3}) reduces
to the classical Stefan problem on the diffusive motion of the interface 
between the phases. In this case, the thickness of the liquid layer is 
expected to grow in time as $H(t)=2 \lambda \sqrt{\kappa t}$,
where the coefficient $\lambda$ depends on the Stefan number according to 
$\mbox{erf}(\lambda) \sqrt{\pi} \lambda St=e^{-\lambda^2}$
\cite{rubinshteuin1971stefan}.

We performed direct numerical simulations of equations
(\ref{eq1}-\ref{eq3}) by a pseudo-spectral code developed for Nvidia 
hardware. The code makes use of OpenACC directives which enable the use
of Nvidia GPU and in particular the CUDA FFT libraries. 
The same code is used for both 2D and 3D simulations and we performed 
runs at the two different resolutions (on uniform grid) 
$1024 \times 1024 \times 512$ and $2048 \times 1024 \times 256$ 
($1024 \times 512$ and $2048 \times 256$ in 2D) 
to check the independence of the results on the domain size. 
Results of 2D and 3D simulations are average over, respectively, $
8$ and $2$ independent realizations to
achieve statistical convergence. 

The boundary conditions at the top and bottom of the domain are enforced
using a penalization term,
while periodic boundary conditions are applied in the horizontal directions.
In all the simulations the fluid is initially at rest, ${\bf u}=0$. 
The initial vertical temperature profile is the linear ramp
described above, connecting the temperature $T_1=4 ^o C$ at the ground surface 
to $T_0=-1 ^o C$ in the permafrost bulk.  
A small random perturbation is added to the initial temperature profile
in order to trigger the linear instability. 
Units for space and time are meters and days, respectively, and
the physical parameters are $H_0=0.33 \, m$, $\kappa=0.012 \, m^2/day$,
$\mu=86.0 \, kg/(m \, day)$, $\varphi=0.5$ and $L=3.3 \times 10^{5} \, J/kg$ 
(for simplicity we use the same values for liquid and ice phase).
We performed three classes of simulations with 
permeabilities $k=4.4 \times 10^{-9} m^2$ (A), $k=5.9 \times 10^{-9} m^2$ 
(B) and $k=8.8 \times 10^{-9} m^2$ (C), typical of water-saturated gravel. 
The temperature jump gives a density difference 
$\Delta \rho=\rho_0 \alpha^{*} T_{max}^q = 0.127 \, kg/m^3$ and
the corresponding characteristic velocities are respectively 
$U_0=0.955 \, m/day$ (A) , $U_0=1.28 \, m/day$ (B) and $U_0=1.91 \, m/day$ (C).
The corresponding initial Rayleigh-Darcy number ($Ra_0$) are
$Ra_0 \simeq 26.3$ (A), $Ra_0 \simeq 35.2$ (A), and $Ra_0 \simeq 52.5$ (C). 

\section{Results and Discussions}

The initial evolution of the model (\ref{eq1}-\ref{eq3}) depends on the 
permeability which determines the value of the initial Rayleigh-Darcy number.
In the cases A and B with $Ra_0 < Ra_c$, the initial diffusive growth of the width of the thawed layer $H(t)$ 
causes an increase of the Rayleigh-Darcy number which eventually reaches the instability threshold,
and the convective motion sets in. 
In the case of run C with $Ra_0 > Ra_c$ convective motion develops from the beginning of the simulation.

\begin{figure}[h!]
\begin{center}
\includegraphics[trim=0 -2cm 0 0,width=0.8\textwidth]{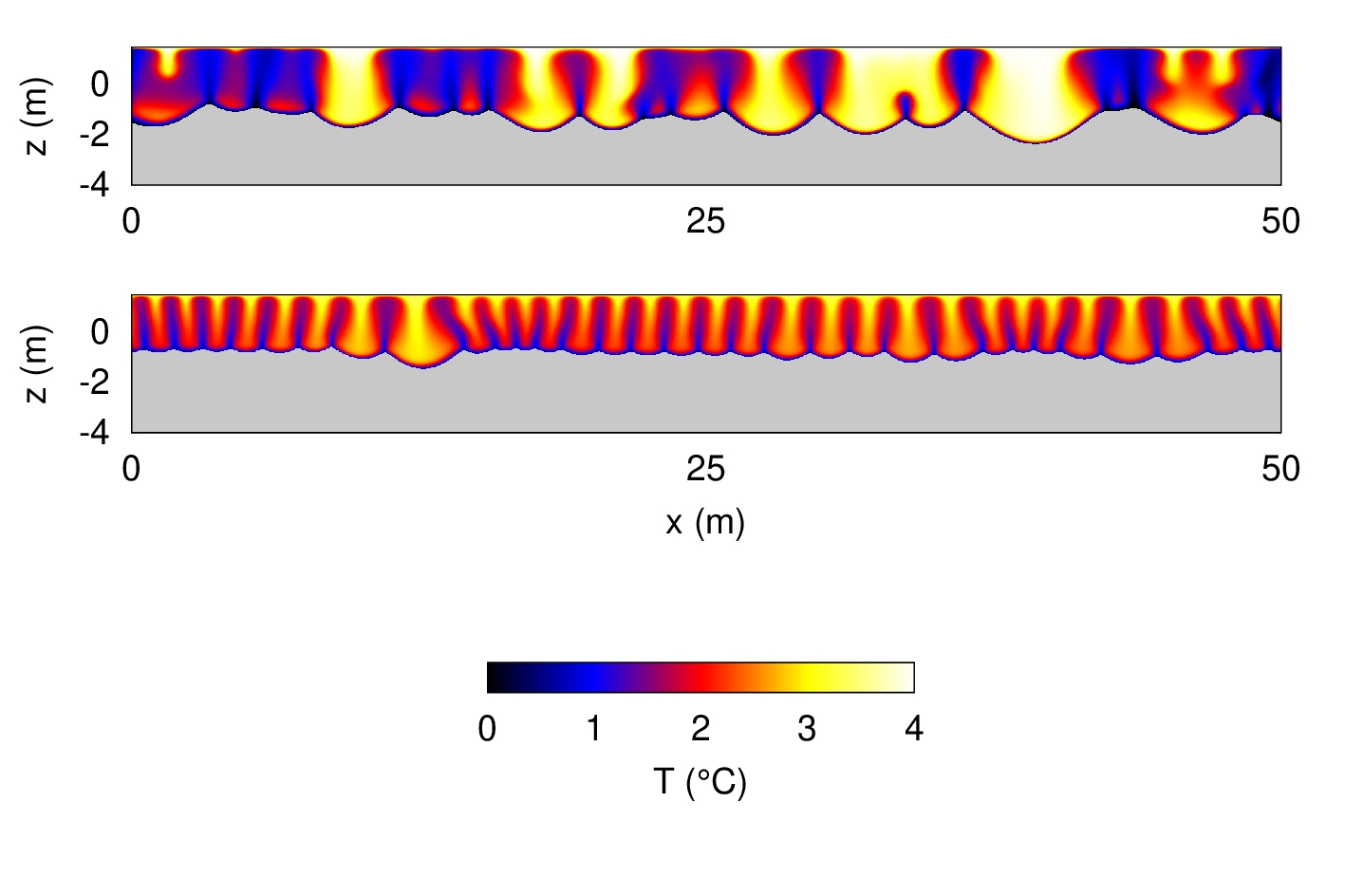}
\end{center}
\caption{
Vertical sections of the temperature field for a 3D simulation (upper
panel) and a 2D simulation (lower panel) of run C at time $t=250 \, days$.
The grey area represents the frozen region where $\phi=1$.}
\label{fig1}
\end{figure}

The typical appearance of the developed convective regime is represented in 
Figure~\ref{fig1}, which displays a vertical section (in the $x-z$ plane at $y=0$)
of the temperature field at time $t=250 \, days$ for a 3D and a 2D simulation with parameters of run C.
The color scale represents the temperature in the thawed region
while the frozen region is represented in grey. 
In both cases, we observe the presence of convective plumes which are 
generated close to the upper layer and move downwards producing an
inhomogeneous heating and thawing of the ice.
As a consequence the fluid-ice interface is not flat,
and ice and water coexist at a given depth.
Already at this level it is evident that the dynamics in the 
2D and 3D cases is qualitatively similar but with important quantitative
differences. In 2D, thermal plumes are narrower and more regular than in 3D case.
Moreover, the average position of the interface is deeper in the 3D case than in 2D at the same time.

In the first part of this section, we present the results
regarding the global evolution of the system averaged over horizontal planes.
In the second part, we investigate the statistics
of the spatial modulation of the fluid-ice interface. 

\begin{figure}[h!]
\begin{center}
\includegraphics[trim=0 -2cm 0 0,width=0.7\textwidth]{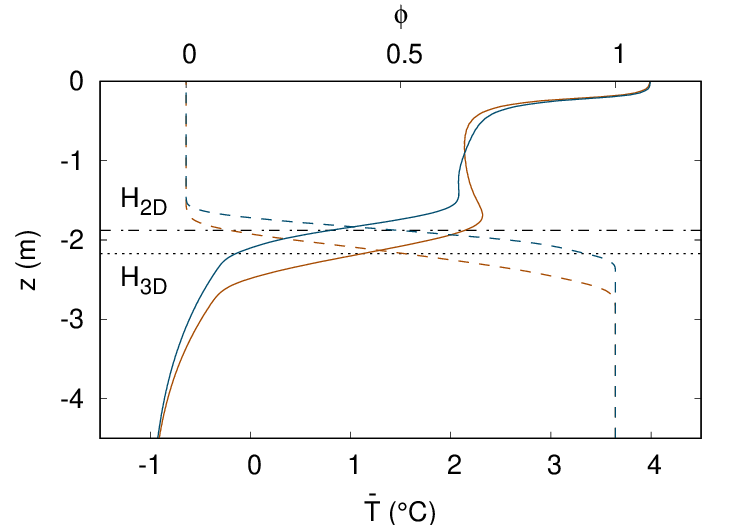}
\end{center}
\caption{
Vertical temperature profile $\overline{T}(z)$ (continuous lines) and
phase field profile $\overline{\phi}(z)$ (dashed lines)
at time $t=150 \, days$ for the 3D simulation (red lines) and 
2D simulation (blue lines) of run C.
The black dotted (dash-dotted) line represents the average position of 
the interface defined as $\overline{\phi}(-H)=1/2$ for the 3D (2D) case.}
\label{fig2}
\end{figure}

\subsection{Thawing dynamics}
In order to characterize the evolution of the system, we consider the
vertical profiles of the temperature field $\overline{T}(z,t)$ and of
the phase field $\overline{\phi}(z,t)$, where the overbar 
represents the average over horizontal planes. 
Figure~\ref{fig2} represents an example of the profiles for 
run C at an intermediate time in the convective regime.
In the temperature profiles we observe the formation of a thin boundary layer
just below the surface level $z=0$
and a local maximum of temperature close to the bottom of the thawed layer.
The latter effect is a signature of the warm tips of the descending thermal plumes.
Below this level, the temperature decreases to the melting temperature $T = 0~^{o}C$,
while the phase-field profile $\overline{\phi}(z,t)$
changes from the liquid value $\phi=0$ to the solid one $\phi=1$.
In the frozen part of the domain (where $\overline{\phi} =1$) the temperature
gradually approaches the bulk value $T_0=-1~^{o}C$. 
The transition of $\overline{\phi}(z,t)$ does not occur sharply at a specific depth,
but rather extends across a region given by $0 < \overline{\phi}(z,t) < 1$,
in which solid and liquid phases coexist at a given depth $z$. 
This is consistent with what has been observed in Fig.~\ref{fig1}. 
Based on this result, we define the average height $H(t)$ of 
the thawed layer by the level at which $\overline{\phi}(-H,t)=1/2$. 

\begin{figure}[h!]
\begin{center}
\includegraphics[width=0.7\textwidth]{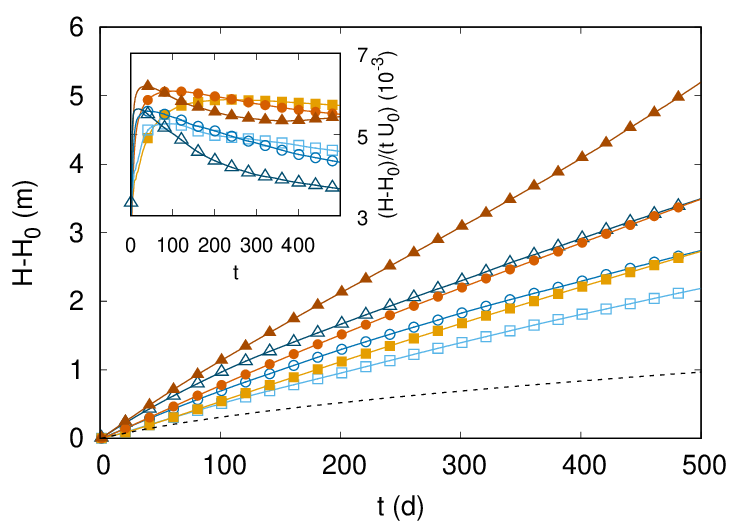}
\end{center}
\caption{Average displacement of the water-ice interface as a function of time 
for the three runs in the convective regime A (squares),
B (circles) and C (triangles) for both 3D (filled symbols, warm colors) 
and 2D (open symbols, cold colors) simulations. 
The black dashed line corresponds to a diffusive simulation 
with $k=5.9 \times 10^{-11} \, m^2$. 
$H_0=0.33 \, m$ is the initial position of the interface at $t=0$.
Inset: interface displacement compensated with the linear growth $U_0 t$.
The constant plateau indicates the final velocity of the interface in 
unit $U_0$.}
\label{fig3}
\end{figure}

The temporal evolution of
$H(t)$ is plotted
in Fig.~\ref{fig3} for both 2D and 3D simulations.
For comparison, we also show the diffusive
behaviour of a simulation with small permeability $k=5.9 \times 10^{-11} m^2$. 
We observe that in all the simulations the thawed layer grows much 
faster than in the diffusive case. 
In agreement with the qualitative picture shown in Fig.~\ref{fig1},
we find that, at fixed permeability, the 3D solution is more efficient
in thawing the underlying frozen layer and the extension of the 
liquid layer grows faster than in 2D. 
At late times, in the 3D simulations
the deepening of the active layer is approximately 
linear in time and proportional to characteristic velocity $U_0$,
in agreement with \cite{magnani2024convection}. 
This is shown in the inset of Fig.~\ref{fig3}, where 
$(H(t)-H_0)/(U_0 t) \simeq 6 \times 10^{-3}$. 
In contrast, the deepening of the active layer does not reach a
constant asymptotic speed in 2D simulations.

\begin{figure}[h!]
\begin{center}
\includegraphics[width=0.7\textwidth]{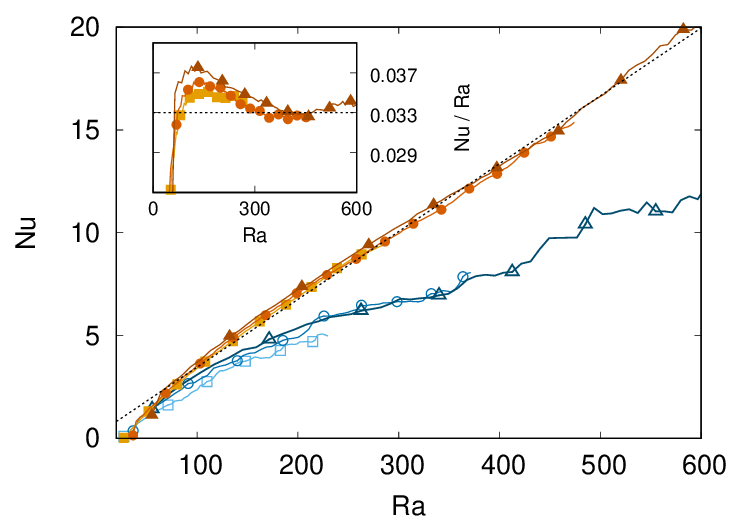}
\end{center}
\caption{Nusselt number as a function of the Rayleigh number for the 
3D simulations (filled symbols, warm colors) and the 2D simulations 
(open symbols, cold colors). Colors are the same as in Fig.~\ref{fig3}. 
Inset: compesated plot $Nu/Ra$ vs $Ra$.
The black dotted lines represents the 
fit $Nu=0.033 Ra$.}
\label{fig4}
\end{figure}

The enhancement of the heat transport due to the convection
with respect to the thermal diffusion
is quantified by the Nusselt number $Nu$,
defined as the ratio of the convective heat flux to the diffusive one

\begin{equation}
Nu = {\langle u_z T \rangle H \over \kappa (T_{max}-T_M)} 
\label{eq5}
\end{equation}
where $\langle \cdots \rangle$ indicates the average over
the convective region of depth $H$ and $u_z$ is the vertical
velocity. 
Figure~\ref{fig4} shows the evolution of the Nusselt number 
as a function of the Rayleigh-Darcy number. 
As the thickness of the active
layer $H$ grows with time, so does the Rayleigh-Darcy number (\ref{eq4}) and 
therefore each simulation spans a range of $Ra$.
Hence, at comparable Rayleigh number values (i.e. comparable active layer depth) 
the heat transport is stronger in 3D, 
with $Nu$ being $\sim 40\%$ larger than in two dimensions. 
This indicates a less efficient transport of temperature by the 2D plumes
which explains the lower speed of the convective process in this case.

At any give time during the evolution, the thawed layer may be thought 
as an analogue of a Rayleigh-Darcy system:  an ideal model consisting 
of a fluid-saturated porous medium confined between 
two rigid boundaries and subject to an unstable density stratification. 
Our results, indeed, agree with the difference observed in Rayleigh-Darcy system
\cite{hewitt2014high,pirozzoli2021towards,depaoli2022strong}. 
In Rayleigh-Darcy, the Nusselt number is lower in 2D than in 3D simulations for $Ra > 200$, 
where the two-dimensional convective modes become subdominant and give 
way to three-dimensional convective plumes. 
In our simulations, given the small density differences of water between $T_M$ and 
$T_{max}$, the Rayleigh-Darcy number remains moderate
for all the investigated permeability values, 
yet the difference of $Nu$ between 2D and 3D runs is clearly visible for $Ra > 200$.  

Figure~\ref{fig4} also shows that, in the 3D case for sufficiently
large values of the Rayleigh-Darcy number 
$Nu$ is approximatively proportional to $Ra$. As explained in 
\cite{magnani2024convection}, this can be undestood by a simple dynamical
argument. From the definition (\ref{eq4}) we have that $Ra$ is 
proportional to the thickness $H$ and, if we assume that the correlation
in (\ref{eq5}) can be approximate as 
$\langle u_z T \rangle \simeq \beta U_0 \Delta T$ ($\beta$ is a 
dimensionless constant), we obtain the simple relation
$Nu = \beta Ra$ which is observed in Fig.~\ref{fig4}.
Remarkably, this relation is independent on the permeability, 
suggesting that the correlation coefficient $\beta$ is independent on the 
properties of the porous medium; yet it depends on the dimensionality of the system. 
We remark that the observed long-term linear $Nu-Ra$ relationship can not be
straightforward interpreted as the achievement of the so-called ultimate regime,
in analogy with Rayleigh-Darcy system.
In the Rayleigh-Darcy system, heat transfer is increasingly well described by a 
heat-exchanger model as $Ra$ increases, for which root mean square (rms) 
temperature and vertical velocity tend to the same value  \cite{hewitt2014high,hewitt2012ultimate}. 
In contrast, in our system the temperature rms measured at the half depth 
of the active layer remains much higher than the vertical velocity rms, even on
long times. 


\subsection{Statistics of the thawing surface}

\begin{figure}[h!]
\begin{center}
\includegraphics[width=0.7\textwidth]{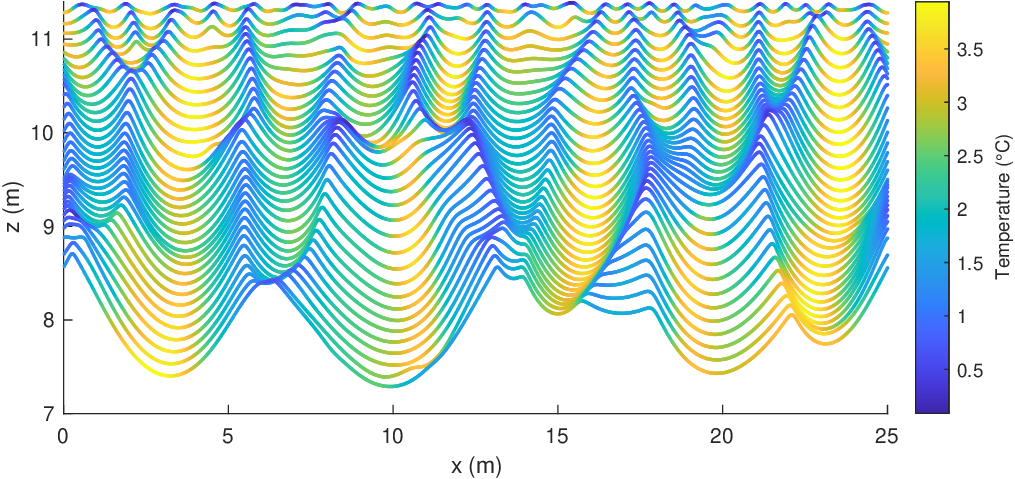}
\includegraphics[width=0.7\textwidth]{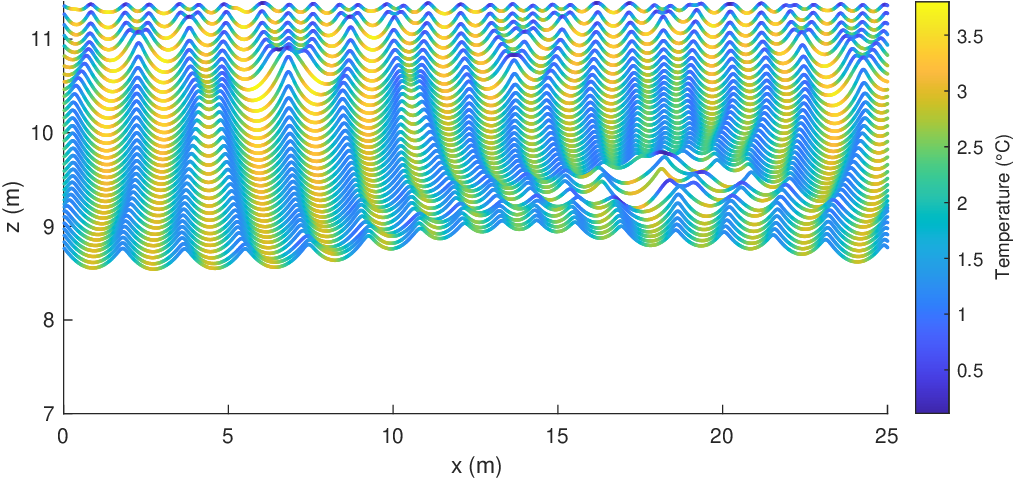}
\end{center}
\caption{Time evolution of the interface position $h(x,t)$ for run C in 3D (top plot) and 2D (bottom plot). 
Colors correspond to temperature $T(x,H/2,t)$ in the middle of the mixing layer,
at $z=H(t)/2$.}
\label{fig5}
\end{figure}

The thawing process is governed not only by the heat transport within the thawed layer
but also by the morphology of the ice-water interface. 
Namely, a larger interfacial area exposed to the convective 
flow is expected to enhance heat exchange and accelerate thawing. 
The coupling between
plumes and the interface is clear from Fig.~\ref{fig1} where 
cold plumes are mainly produced at the tips of the interface. 
On the contrary, warm descending plumes concentrate in the troughs
of the interface and this process produces even deeper 
regions (see also Fig.~\ref{fig5}). Hence, the shape of the thawing interface reflects 
the underlying convective flow structures while simultaneously shaping the flow itself. 
This feedback raises the question of how the interface evolves over time under 
the combined influence of convection and its own morphological impact on the flow.

Figure~\ref{fig5} shows the evolution of the 
interface section $h(x,t)$ (defined as the points at which the phase field 
$\phi(x,y=0,h,t)=1/2$) for the run C at largest permeability. 
For both 2D and 3D simulations, we observe that troughs become 
deeper during the evolution because of the enhanced melting below warm downward plumes. 
From Fig.~\ref{fig5} it is also evident that troughs not only deepen but also widen owing to 
the merging between nearby troughs (and the associate convective
plumes) and the disappearance of the internal tips. 
This morphological evolution is markedly faster in 3D than in 2D and produces much
wider troughs at the final time. 

In order to characterize quantitatively the interface profile, we
define the characteristic trough width 
$\Delta x(t)$ as the average separation between two tips 
(local maxima) on the interface. Similarly, the characteristic height 
$\Delta z(t)$ of the interface is computed as the average vertical distance 
between one local maximum (tip) and the following minimum (trough). 

\begin{figure}[h!]
\begin{center}
\includegraphics[width=0.7\textwidth]{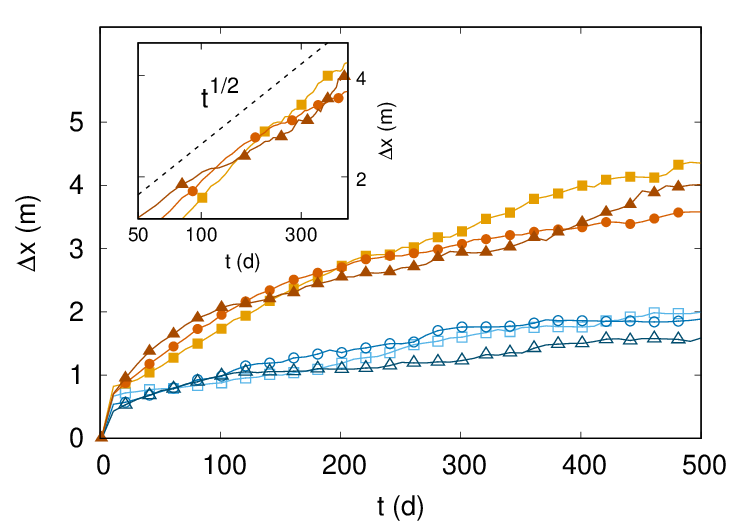}
\end{center}
\caption{Average trough width $\Delta x$ on the thawing interface as a 
function of time for the 3D simulations (filled symbols, warm colors)
and 2D simulations (open symbols, cold colors). The inset shows the
3D results in log-log plot to compare with the $t^{1/2}$ growth. 
}
\label{fig6}
\end{figure}

Figure~\ref{fig6} plots the time evolution of the average trough width
$\Delta x$ for the three runs in both 2D and 3D simulations. 
In all cases we find that $\Delta x$ weakly depends on the permeability,
but we have a substantial effect of dimensionality. 
In the 3D case, the time evolution is compatible with diffusive behavior,
$\Delta x \propto t^{1/2}$, as shown in the log-plot in the inset. 
Remarkably, this is reminiscent of a coarsening process in two (or more)
dimensions of a non-conserved order parameter \cite{bray2003coarsening}.
Indeed, in this case the coarsening process occours on the 2D interface
$h(x,y)$ of which the sections of Fig.~\ref{fig5} are one-dimensional
cuts. 
Although the dynamics in this case is not driven by surface tension,
the process of merging of thermal plumes (and of patterns on the 
interface) apparently follows the same scaling law.
The interpretation of interface evolution in terms of coarsening
process is also supported by the dependence on the dimensionality of our
results. In 2D convective simulations the interface is a 1D
line, a case in which the coarsening process is known to be
much slower, as shown in Fig.~\ref{fig6}.  

We also remark here that the the widening of troughs and plumes and 
its dependence on the dimensionality appears to be a consequence of 
the interaction between thermal plumes and the dynamical thawing interface.
Indeed in the absence of an interface, numerical simulations of
porous convection in the Darcy-Rayleigh-Taylor configuration show that
the size of plumes grows following a diffusive law independently on the
dimensionality \cite{borgnino2021dimensional}. 
In the case of Rayleigh-Darcy convection between fixed and flat boundaries,
numerical studies have shown that the typical size of the structure
shrinks by increasing the Rayleigh number 
\cite{hewitt2012ultimate,hewitt2014high}, a behavior opposite to what
observed here. 


\begin{figure}[h!]
\begin{center}
\includegraphics[width=0.7\textwidth]{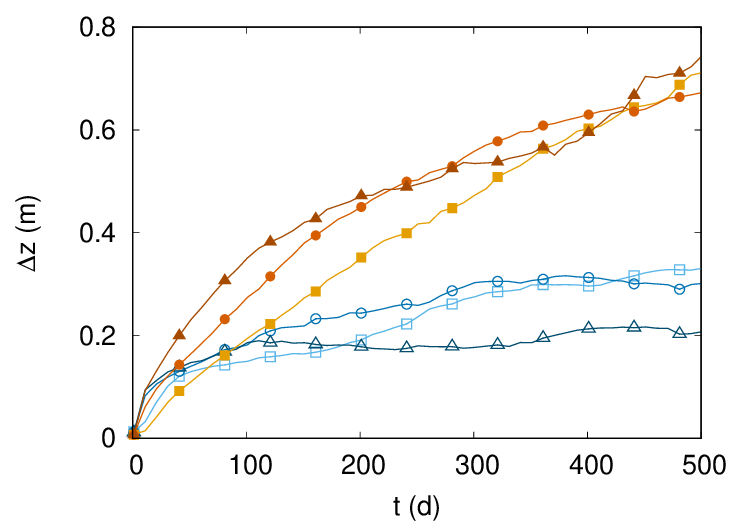}
\end{center}
\caption{Average trough height $\Delta z$ on the thawing interface as a 
function of time for the 3D simulations (filled symbols, warm colors)
and 2D simulations (open symbols, cold colors). 
The inset shows the relationship between $\Delta x$ and $\Delta z$.
}
\label{fig7}
\end{figure}

The evolution of the characteristic height of the interface $\Delta z(t)$
is shown in Fig.~\ref{fig7}. We observe a behavior similar to $\Delta x(t)$
for both 2D and 3D simulations, and this indicates that the shape of the
interface is simply rescaled in time during the evolution. 
In the 3D case the growth of $\Delta z(t)$ is indeed compatible with 
a diffusive behavior. 
It is worth to remark that an analogous growth of the interface roughness
has been recently investigated in turbulent bulk flow where a scaling 
with the Rayleigh number has been observed \cite{yang2023morphology}.

We note that the progressive deepening of interface troughs 
locally enhances the Rayleigh number, 
thereby promoting a stronger convective heat transport (Fig. \ref{fig4}). 
This mechanism underlies the feedback described above, 
whereby interface morphology and convective plumes reinforce each other. 
In addition, the simultaneous increase in trough depth and width 
indicates a progressive enlargement of the ice–water interfacial area 
available for heat exchange. This enlargement is consistently 
higher in 3D than in 2D simulations. Together, these two mechanisms 
explain the faster thawing rate observed in Fig. \ref{fig3}.

\section{Concluding remarks}

We have studied ice thawing in porous media by means of 
direct comparison of 2D and 3D numerical simulation within the 
incompressible Darcy approximation.
In the presence of convection we observe a complex topography of the 
ice-water interface emerging from the initial flat surface and
characterized by a modulation with growing width and depth.
The evolution of the characteristic wavelength of the interface 
in 3D follows a diffusive law, reminiscent of a coarsening process,
while it appears to be slower in 2D. 

The interface morphology affects the position of thermal plumes in the 
above thawed layer thus producing a feedback on convection which 
eventually affects the global displacement of the interface itself.
As a consequence of this feedback, the velocity of the ice-water interface 
is slower in 2D than in 3D. This is remarkably in contrast with convection in 
porous media in the absence of phase transition, e.g. in the 
Rayleigh-Taylor-Darcy configuration, where the opposite behavior 
is observed, i.e. faster plumes in 2D geometry than in 3D
\cite{borgnino2021dimensional}.

Another important observation is that the Nusselt number
in 3D simulations is larger to that in 2D simulation at the same
Rayleigh number, indicating the importance of the three-dimensional
structure of thermal plumes for the transport of the temperature field. 
Remarkably, we observe that in 3D for sufficiently large values of the 
Rayleigh numbers, $Nu$ becomes linearly proportional to $Ra$, independently
on the permeability of the porous medium.


It is important to recall that, large permeability values are needed to
reach the condition for the fluid density stratification to become 
unstable and undergo convective motions. Such high permeability values
are for instance met in rocky and sandy-gravel soils, corresponding to 
the range of parameter values used in the present study 
\cite{gleeson2011,chapuis2012,ebel2019}. 
Under these conditions, climate models accounting for only heat conduction
may underestimate the permafrost degradation. 
Despite using a simplified representation of the active layer system, 
the framework proposed here can serve to obtain useful parameterisations
to be implemented in climate models. Yet, we showed that using two-dimensional 
simulations instead of three-dimensional ones may lead to underestimate 
permafrost degradation.

{\it Acknowledgements.} 
We acknowledge the financial support from Fondazione CRT for the project
"DisPerm - Dinamica del disgelo del permafrost" and HPC CINECA for computing 
resources (INFN-CINECA Grant INFN26-FieldTurb).

\bibliography{biblio}

\end{document}